\documentclass{article}
\usepackage{cite}
\usepackage{amsmath,amssymb,amsfonts}
\usepackage{algorithmic}
\usepackage{graphicx,color}
\usepackage{float} 
\usepackage{textcomp}
\def\BibTeX{{\rm B\kern-.05em{\sc i\kern-.025em b}\kern-.08em
    T\kern-.1667em\lower.7ex\hbox{E}\kern-.125emX}}
\AtBeginDocument{\definecolor{ojcolor}{cmyk}{0.93,0.59,0.15,0.02}}
\begin{document}

\title{
Tunable Experimental Testbed for Evaluating Load Coordination Methods}

\author{Drew A. Geller and Johanna L. Mathieu\footnote{Geller is with Los Alamos National Laboratory, Los Alamos, NM, 87544 USA, dgeller@lanl.gov. Mathieu is with the Department of Electrical Engineering and Computer Science, University of Michigan, Ann Arbor, MI, 48109 USA, jlmath@umich.edu. This research was supported by the U.S. Department of Energy's Advanced Research Projects Agency - Energy (ARPA-E), grant number DE-AR0001061.}}

\maketitle

\section*{ABSTRACT}
Driven by the need to offset the variability of wind and solar generation on the electrical grid, development of load controls is a highly active field in the engineering literature.  However, practical use of residential loads for grid balancing and ancillary services remains rare, in part due to the relative cost of communicating with hordes of small loads and also due to the limited experimentation done so far to demonstrate reliable operation.  To establish a basis for the safe and reliable use of fleets of small compressor loads as distributed energy resources (DERs), we have constructed an experimental testbed in a laboratory, so that load coordination schemes can be tested at extreme conditions within a laboratory environment.  This experiment can be used to tune a simulation testbed to which it can then be linked, thereby augmenting the effective size of the ensemble of loads.  Control algorithms can simply be plugged in for testing.  Modeling of the system was done both to demonstrate the experimental testbed's behavior and also to understand how to tune the behavior of each participating model house in the system.  Implementing this testbed has been useful for the rapid turnaround of experiments on various control types, and it enables testing year-round without the constraints and limitations arising in seasonal field tests with real human participants.
\section{INTRODUCTION}
New deployments of renewable energy resources have increased rapidly over the last decade, driven both by 
new policies to address climate change and by economics as the price of renewable generation rapidly decreases~\cite{cali}.  Because the fastest growing forms of renewable generation -- solar and wind -- are variable and intermittent, more fast-response storage and load control capacity will be needed to maintain a reliable grid as the generation mix evolves with a reduced fraction of schedulable/controllable generation resources.  Load coordination, however, is difficult because different types of electrical loads have different characteristics and customer requirements, so that the reliability, consistency, and performance
of aggregations of loads is uncertain.  For load aggregations to gain acceptance as grid balancing resources and avoid liability with both customers and independent system operators (ISOs), it is necessary to test and debug control schemes to failure in an experimental environment before moving to field testing with real customers.

In this paper we consider aggregations of air conditioners (ACs), which are thermostatically controlled loads (TCLs) where the power consumption can be controlled by sending commands or temperature settings directly to the thermostats.  One reason for selecting ACs is because of their ubiquity, with ACs present in almost 90\% of single-family homes in the U.S.~\cite{recs}.  Another reason is that residential air conditioners consume more energy than other home appliances such as refrigerators (also a TCL), so they have a larger impact on the electricity supply-demand balance if their consumption can be shifted in time~\cite{callaway_tapping_2009}.  ACs and other compressor-based loads are more challenging to control than resistive loads like electric heaters, because compressors have a ``lockout" time between when they are shut off and can be turned on again.  Short-cycling the device risks causing the compressor to stall and draw a high current until it overheats or trips a circuit breaker.  ACs also display a variable power draw as they start up and reach equilibrium temperatures on the hot and cold heat exchangers.  Finally, the AC's compressor draws a transient inrush current when switched on, like any motor, and the impact on transformers and distribution circuits of multiple ACs switching on at exactly the same time should be considered.

We constructed such an experimental testbed at Los Alamos National Laboratory to study the effects of controls on residential window air-conditioners in nearly constant ambient conditions.
The testbed consists of twenty single-zone model houses of nearly identical construction, situated within a warehouse.  
To ensure that these models were representative of real homes of utility customers, the experiment was run open-loop with a simple bang-bang thermostat and validated against field data supplied by Pecan Street, Inc. for 47 homes in Austin, TX~\cite{PSI}. 
Because even twenty ACs represents only a small aggregation, the data acquisition software is designed to optionally interact with a simulation of hundreds or thousands of virtual houses to boost the effective size of the testbed, so that controls may be implemented on a more realistic number of ACs.
Confidence in these integrated experiments can be established by comparing results from the ``real" model houses to the virtual models, to check that the virtual houses match the ``real" ones and that the control algorithm does not favor either set for responding to a regulation signal.
Additional features of the testbed include a programmable heat source, to emulate changes in occupancy of the house, and the ability to tune the natural cycling period of ACs in the house with minor physical modifications.


The experimental testbed described here is one of a very small number of laboratory experiments that have been built at this scale, e.g., \cite{bindner2016,supermarket,vrettos_experimental_2018a, vrettos_experimental_2018b}.  In contrast to our testbed, \cite{bindner2016} focused on refrigerators; specifically, the authors investigate the ability of refrigerators to provide secondary frequency regulation via experiments on 25 refrigerators. Refs.~\cite{supermarket,vrettos_experimental_2018a, vrettos_experimental_2018b} focused on commercial rather than residential buildings; \cite{supermarket} explores demand response from a commercial refrigeration system; and \cite{vrettos_experimental_2018a, vrettos_experimental_2018b} used an experimental facility called FLEXLAB~\cite{flexlab} to demonstrate frequency regulation with a commercial HVAC system. 

In Section \ref{testbed-design} we describe the features and design of our experimental testbed.  Section \ref{testbed-behavior} describes some of the observed features of the testbed during operation with an on-off thermostat emulated in software. Some of this behavior, such as the time-varying power consumed by the ACs when on, may be neglected in simulations but should be considered for accuracy. Section \ref{ETPmodeling} describes our extensions to the equivalent thermal parameter (ETP) model~\cite{etp-model}, and it shows how this model was used to understand and tune the thermal cycling durations of the model houses. Section \ref{all-testbed} summarizes the operation of the 20-node testbed. Finally, the paper concludes in Section \ref{conclusion-section}.  In the Appendix, we provide the simple AC model used in our extension of the ETP model.

In summary, the contributions of our paper are as follows. 1) We present the design and validation of an experimental testbed for testing aggregate AC load control strategies; 2) we develop an extended ETP model of ACs, which captures more of the salient characteristics of these devices; and 3) we highlight the opportunities and challenges associated with testing AC load control strategies through physical experiments. These findings may be of interest to researchers and practitioners exploring methods to test novel load control strategies and test/benchmark production-ready strategies before pushing them to the field. These findings also highlight some of the inherent challenges in using aggregations of ACs for grid balancing services, and some techniques for overcoming those challenges.  

\section{Testbed Design} \label{testbed-design}
Although some houses have two or more zones, the testbed model houses are single-zone units.  The model consists of the air conditioner, a heat source, a circulation fan, and an insulating shell.  Although the models are not exactly scaled from any particular house, they exhibit salient features found in typical field test houses~\cite{PSI}, as will be seen below.  Limiting the models to single-zone systems is not especially restrictive, as many single-family houses have single-zone central air conditioning. Also, individual window air conditioners are found widely in single rooms of houses, in apartments, in hotel rooms, and in small offices.  

\begin{figure}
\centering
\includegraphics[width=8.5 cm]{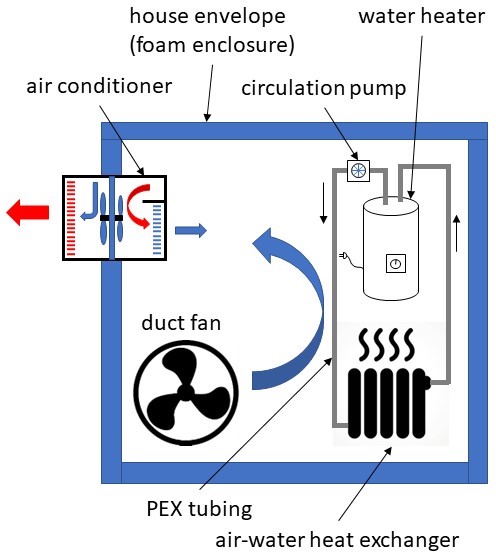}
\caption{Model houses are constructed inside a four foot cubic foam box.  The internal heat source consists of a hydronic loop with an electric water heater.  A duct fan forces air through the heat exchanger and mixes the room air in addition to the AC's fan.\label{model-house}}
\end{figure}

The model houses may be broken into subsystems: (1) the air conditioner, which is the TCL to be controlled, (2) the enclosure and environment, (3) the thermal properties including heat sources, and (4) the instrumentation and control system.  Each of these is considered in the subsections below, and a sketch depicting this arrangement is given in Fig.~\ref{model-house}.

\subsection{Air Conditioners}
The ACs used in this testbed are single-speed.  Although dual-speed and variable-speed air conditioners are more efficient and provide improved user comfort, they are not yet dominant in the U.S.  Single-speed ACs are cheaper and widely available, and for central air systems they do not require improved insulation around ducts to avoid condensation.

To accommodate as many TCLs as possible, we selected 5000 BTU/h window air conditioners, which are the smallest capacity commonly sold.  The specific model selected contained a mechanical thermostat, and the thermostat could be easily replaced with a relay controlled externally by the data acquisition computer.  Also, the ACs have a single speed fan, which was turned ON all the time both to provide additional air circulation in the box and also to distinguish its power consumption from that of the compressor.  The air conditioners are equipped with a passive element that interrupts power if the compressor stalls and overheats, but adjustable switching delays or lockout periods were imposed in the data acquisition software to prevent short-cycling the compressor.  Because the built-in mechanical thermostat was not used, the computer control provided the thermostat function in software, switching based on temperatures measured by negative temperature coefficient (NTC) thermistors in each house.  These thermistors were specified by the manufacturer as directly replaceable with an accuracy of $\pm 0.2$\%.  Each thermistor was measured in a voltage divider arrangement using precision resistors, with a driving voltage low enough not to affect the temperature readings due to self heating.

\subsection{House Envelope}
The insulating boundary of each house was a foam enclosure of dimensions approximately 4 ft.\ $\times$ 4 ft.\ $\times$ 4 ft.\ , constructed from 2 in.\ thick extruded polystyrene foam boards with an R-value of 10.  The volume is less than 1/12 that of a small, 100 sq.\ ft.\ room with 8 ft.\ ceilings, and so the heat capacity of the air will be correspondingly smaller.  The humidity of the air is normally quite low in Los Alamos, New Mexico, with values around 10\% relative humidity (RH) as typical.  Relative humidity in the lab was measured with a sensor, but the humidity was not controlled in the lab; and for these experiments no adjustments to the data were either needed or attempted to account for humidity.

\subsection{Heat Capacity and Heat Injection}
In real houses, the solid materials exchange heat with the air as well and significantly affect the temperature dynamics through their heat capacities.  The model houses contain electric point-of-use water heaters, of either 20 or 30 gallon capacity, that are intended to represent the solid heat capacity normally comprised by the solid walls and furnishings.  Furthermore, by adjusting the heat exchange between water and air, one may effectively make the model house appear to have a higher heat capacity for the air and thus to cycle the compressor on and off less frequently.  In this way, the small model system can appear to the electrical feeder more like a real house, at least in terms of its cycle durations.  To transfer heat from water to air, there is an approximately 1 ft.\ $\times$ 1 ft.\ tube-and-fin heat exchanger, and water is circulated from the water heater through cross-linked polyethylene (PEX) tubing and a pump to the heat exchanger and back to the water tank.  This is similar to the hydronic loop of a residential radiant heat system.  A single-speed duct fan is placed next to the water-air heat exchanger to flow air through it and mix the air throughout the small box.  

The water circulation pump and the duct fan are fixed electrical loads enclosed in the house, and they run constantly during experiments.  They are therefore constant internal heat loads.  In contrast, the water heater is used as a programmable heat source, as the amount of heat injected is adjusted by a solid-state power controller.  The controller output is proportional to a 4-20 mA signal, which can be generated by an analog voltage from the data acquisition system using a circuit based on a voltage-to-current converter like the Texas Instruments XTR110.  The voltage driving the water heater is stepped down using a power transformer to deliver a maximum heat gain to the house where the temperature would be nearly constant for the AC running continuously. 
This limiting of heat gain is done both for safety and to ensure that the power controller settings span its full-scale range for maximum accuracy in the experiments.  Each water heater has its own power controller circuit, so in principle the houses could all have different internal heat gains.  The data acquisition computer controls all the heating rates and it also allows the operator to add randomness to the heating rate (if needed to disturb phase-locking or synchronization of houses) or to make the heating rate vary over time to emulate occupant usage patterns or diurnal usage variation.

\begin{figure}
\centering
\includegraphics[width=8.5 cm]{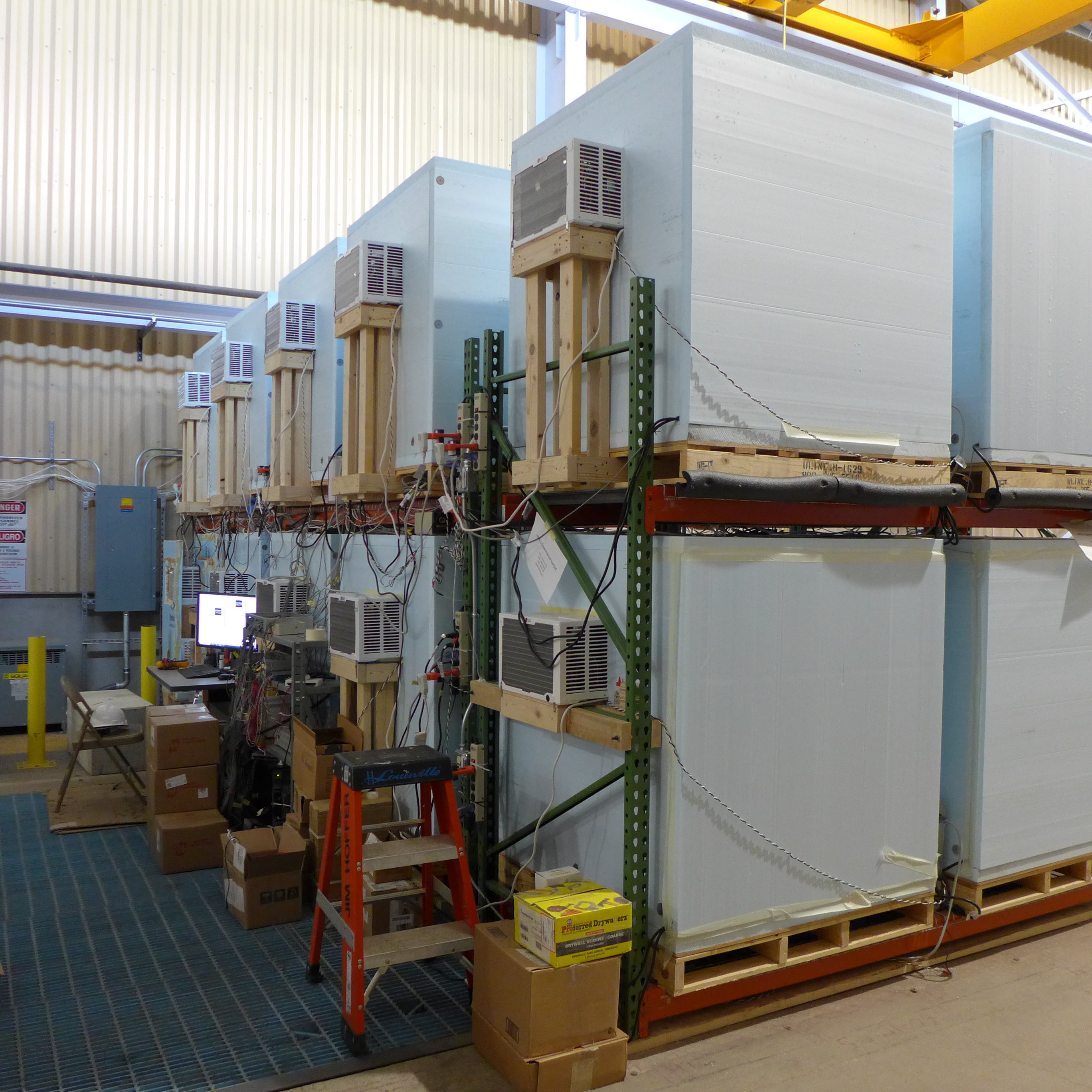}
\caption{The 20 units are stacked in pallet racks, minimizing cable lengths and overall footprint.  In spite of their close proximity to one another, the ACs are not found to synchronize.\label{lab-photo}}
\end{figure}

Limited floor space was available, but the laboratory had a high ceiling and a crane, so the assembly was constructed in the compact, two-level arrangement shown in Fig.~\ref{lab-photo}.  Although most of the houses are assembled with the same components and plan, there was found to be some variability in their natural cycling patterns in the lab so that they constitute a heterogeneous set.  The power usage of the pumps, fans, and air conditioners were found generally to differ by no more than 5\% for identical components, so this was unlikely to be a major cause of variation. 
Several other causes for the inhomogeneity were therefore considered.  First, the water heaters were chosen to be in two different sizes to ensure that there could be some intentional inhomogeneity and the ability for the ensemble to be adjusted to span a wide range of natural cycling periods. Second, small differences in the heat leak through the foam boxes can occur due to the different locations of the units in the testbed.  Some units are close to a north-facing wall of the building and see cooler temperatures than the other units most of the time.  There are temperature gradients in the room, and the elevated units in the two-story structure may occasionally be warmed by sunlight from the warehouse's clerestory windows.  A third source of inhomogeneity may also come from minor differences in the positioning of components and the air flow around them inside the house.    
Finally, the position of the thermistor measuring the air temperature near the air-water heat exchanger makes the most difference because of the temperature of air flowing past it, as will be seen in the discussion below.

\subsection{Data Acquisition and Control}
The entire array of model houses, the laboratory sensors, and the power meter monitoring of each house are managed by a single computer system running a  data acquisition program written in NI LabView.
Most of the signals, such as thermistor readings, power levels to the water heaters, and the digital ON/OFF signals for the compressors, are served by a NI PXIe chassis because of the accuracy, speed, and high channel density of the PXIe interface.  The power meter is an eGauge Pro EG4130 with twenty 100~A current transformers (CTs) attached, one for each house.  The real and apparent power of each CT, plus the voltages and frequencies of the three 120 V circuits in the panel, are read at the meter's maximum rate of 1 Hz over an RS485 connection.  The LabView software includes a simple TCP server that allows it to communicate via the loopback network with other programs on the same computer.  In early experiments, for example, MATLAB scripts were used to request timed switching patterns for groups of TCLs in the assembly.  The LabView server would switch compressors ON or OFF depending on whether this was consistent with the thermostat configuration loaded into the LabView program at the outset.  In later experiments (to be described in a separate paper), the TCP interface was used to connect to a controller written in MATLAB.  The MATLAB program also contained a simulation of a population of virtual houses to group with the physical model houses for testing control schemes.  In this case, the LabView software both adjudicated control requests against its internal thermostats and transferred data back to the MATLAB controller so that the controller could know the full state information of the experiment's TCLs.  Since the TCL states for the ``real" houses were measured in real time, the clock of the LabView experiment determined the latching of time steps, such that the MATLAB controller would step along with the physical experiment.

\section{Testbed fixed-setpoint behavior} \label{testbed-behavior}
\begin{figure}
\centering
\includegraphics[width=8.5 cm]{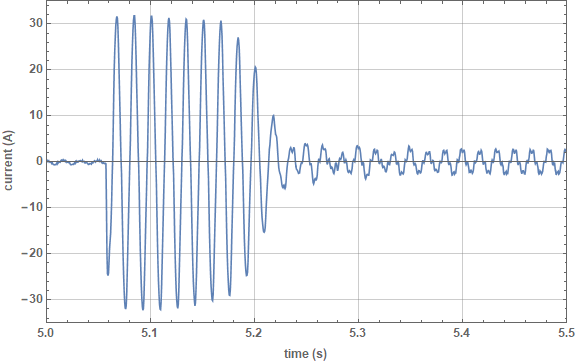}
\caption{Inrush current measured for a single air conditioner.\label{fig:inrush}}
\end{figure}

It is useful to run experiments on physical model houses and locate phenomena that might be overlooked in simple mathematical models.  Such phenomena may cause a control algorithm that worked well in simulation to not work as expected on real systems.  For example, ACs do not simply turn on and have a flat power draw, as they are commonly modeled.  Rather, there are fast and slow features in the power consumption.  Initially, there is a spike in power due to the inrush current, because the compressor takes time to start from rest, where the stalled input impedance is quite small.  It generally takes 5-10 cycles for this inrush current to decay, as seen in Fig.\ \ref{fig:inrush}.  The power associated with the inrush current is a negligible contribution to the power consumption of the air conditioner, averaged over a complete ON/OFF cycle which is generally no less than 4 minutes long.  The inrush current may matter, though, to upstream components in the distribution network, because many ACs turning on simultaneously will briefly draw five or six times the steady-state current expected; the effect of this transient on the network must be evaluated.  

\begin{figure*}
\centering
\includegraphics[width=\columnwidth]{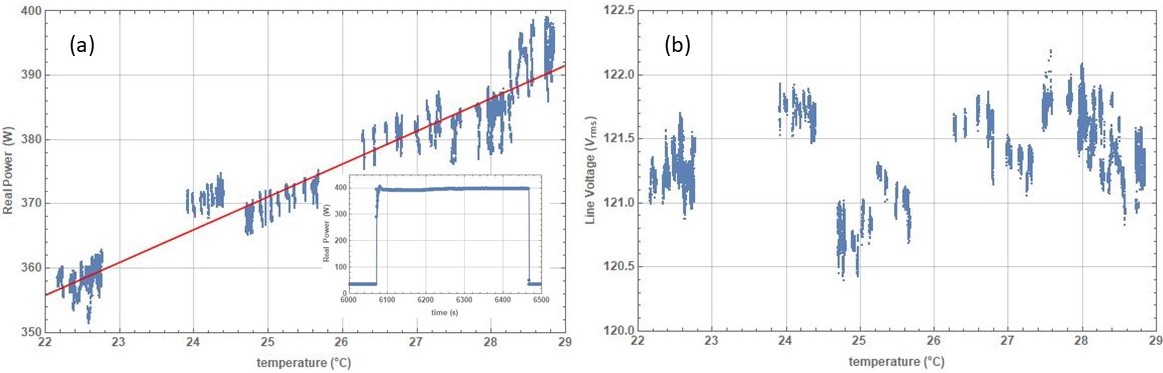}
\caption{(a) Peak power for the air conditioner varies with ambient temperature.  Here, data from a single air conditioner is compiled for runs under identical heat injection but varying lab temperatures.  The data shown are the power during each ON-state, dropping the earliest data after the compressor starts.  Because the power consumption continues to rise slowly and does not reach its asymptotic limit before the thermostat shuts it off, each cycle contributes a short segment of points instead of a single point at its maximum. This and the inrush transient are seen in the inset at the bottom right, depicting a single power ON cycle of the AC. (b)  In contrast, the line voltage is uncorrelated with ambient temperature in the lab.  In field data, the line voltage may droop at higher temperatures as more consumers use their air conditioners and increase the load on the feeder \cite{PSI}--the population of air conditioners drawing from the feeder is not constant.  In the lab, though, the testbed 
always has a fixed population of 20 air conditioners operating, so no voltage droop is seen.  Each air conditioner will draw slightly more power at higher ambient temperatures, but the incremental change in aggregate power consumption compared to all loads on our transformer is too small to detect.
\label{fig:pVsT}}
\end{figure*}

Following the inrush transient, the power consumption briefly drops as the compressor delivers refrigerant vapor up to the expansion (or throttle) valve leading to the evaporator.  The power then rises rapidly for a few seconds and with the pressure at the throttle valve, as the vapor is pressurized and condenses to a liquid there.  Then the power begins to rise more slowly and may reach a plateau as the evaporator and condenser heat exchangers approach their minimum and maximum temperatures, respectively.  This evolution in power consumption is shown in the inset to Fig.~\ref{fig:pVsT}(a). The entropy generation in moving heat from low to high temperature requires work from the compressor.  Other power is consumed by friction in the compressor and by the work done in compressing the gas into the condenser.  There are no corresponding delays in power consumption approaching zero when turning off the compressor.

The ambient temperature outside the model houses affects the dynamics of the houses in two ways.  First, there is the heat leak through the walls of the house, which depends on the temperature gradient across the foam wall.  Higher temperatures outside the box will show up as higher duty cycles for the air conditioners, as they need to remove extra heat in each cycle.  In these model houses, the additional heat load was approximately 5 W/$^{\circ}$C through the walls.  Second, even if the heat leak into the box were negligible, the hot heat exchanger must now reject the same amount of heat into a higher-temperature ambient reservoir.  Since the fan's flow rate is fixed and the heat transfer coefficient is unchanged, the hot heat exchanger would increase in temperature to maintain a constant temperature difference $\Delta T$ with the environment.  For these model houses, the increase in power was $1.36\;  \rm{\%/{^\circ}C}$ as shown in Fig. \ref{fig:pVsT}.  This is similar, but not identical, to the fractional change measured in~\cite{CIEE-report} and also to that described by Pecan Street, Inc.\ \cite{PSI} for houses representative of their field testbed for this project.

\begin{figure*}
\centering
\includegraphics[width=\columnwidth]{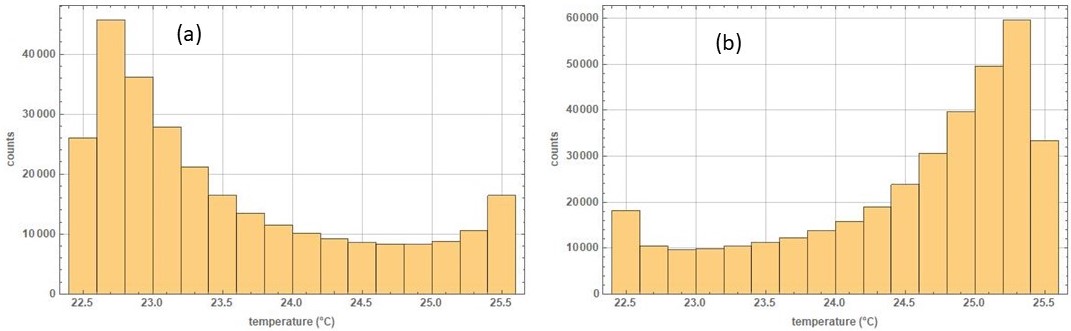}
\caption{Histograms of temperatures for all 20 houses, measured every second and integrated over several hours, (a) while ACs have their compressors ON and (b) while ACs have their compressors OFF.  The ACs are not synchronous in their ON and OFF states.  Rather, each count represents a single house being at a particular temperature when measured in a given 1-second long time step in the experiment.  The internal heating rate for each model house was programmed to be 250 W plus about 125 W for the fixed heat generation of the pump and fan in the unit.  This requires about a 30-40\% duty cycle for the ACs at this setpoint.\label{non-unif-T}}
\end{figure*}

For any of the houses, the 
instantaneous probability of finding the house at any temperature within the temperature deadband over time is not uniform, with the houses spending most of their time near the edges of the deadband.  This is shown in Fig.~\ref{non-unif-T} for all the ACs when they are in the ON state or the OFF state, respectively; and it
occurs because of the non-linear
temperature evolution with time when heat is driven by temperature gradients from the environment (e.g., ambient temperature or solar irradiance) or from heated solids within the house.  The mean temperature, referenced to the set point temperature, will vary almost inversely with the duty cycle of the air conditioner, within the deadband limits.  The duty cycle is defined as
\begin{equation}
    \rm{duty \: cycle} = \frac{\rm{duration \: compressor \: ON}}{\rm{duration \: of \:complete \: cycle}} \times 100 \%.
\end{equation}
For example, if the injected heat is high enough that the AC is on 90\% of the time to make the thermostat reach the lower deadband limit in temperature, then the room must be cooling very slowly as it approaches this limit and the time-averaged temperature must be close to the lower deadband limit $T_-$.  If the injected heat is minimal but the room still warms when the AC is off, then the temperature approaches the upper deadband limit $T_+$ slowly and the average temperature is close to $T_+.$


\begin{figure}
\centering
\includegraphics[width=8.5 cm]{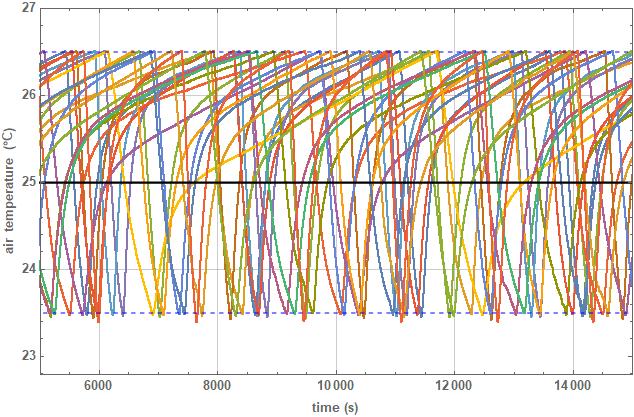}
\caption{Temperature evolution of the model houses with the minimum injected heat.  The thermostat temperatures undershoot the deadband limit by as much as 0.1 $^{\circ}$C.\label{temp-delay}}
\end{figure}

The temperature generally shows a delayed response to turning ON/OFF the compressor, so the temperature exhibits small excursions outside the deadband when the AC changes state.  The air conditioner itself has some thermal inertia in that the compressor must first start to cool the heat capacity of the metal evaporator before the air begins to cool; and, similarly, when the compressor halts, the condensed refrigerant continues to cool the heat exchanger briefly as the condensate vaporizes (Fig.~\ref{temp-delay}).  There is also a time constant for the mixing of the air in the room, limited by the flow rate of the circulation fans in the space ($\approx$ 12 s in these model houses).  Finally, the temperature lag will depend on the location of the room's thermometer, which may be influenced by its proximity or attachment to nearby solid features and which depends on the heat gain in the room.  This ``temperature delay" \cite{temp-delays} could cause  tracking excursions in the controller, since the controller may rely on a prediction of instantaneous thermometer response to the compressor after a state change.  These features are not normally represented in ETP modeling \cite{etp-model}, because the lumped-parameters of the model are a simplification of the real system.

\section{Numerical Model}\label{ETPmodeling}
An equivalent thermal parameter model~\cite{etp-model} was developed for understanding the main features of the individual model houses.  The parameters of the physics-based model were chosen to approximately represent the houses, but the model was not 
calibrated using system identification to represent any of the particular houses.  Rather, the model was used to explore some of the basic behavior with respect to the air-water heat transfer and to the air conditioner itself, to better understand the operation of the houses.  To accomplish this, the ETP model was extended by using a simple model of the air conditioner and by introducing a temperature offset to the thermostat measurement.  These new model features are described in the discussion below.

\begin{figure}
\centering
\includegraphics[width=8.5 cm]{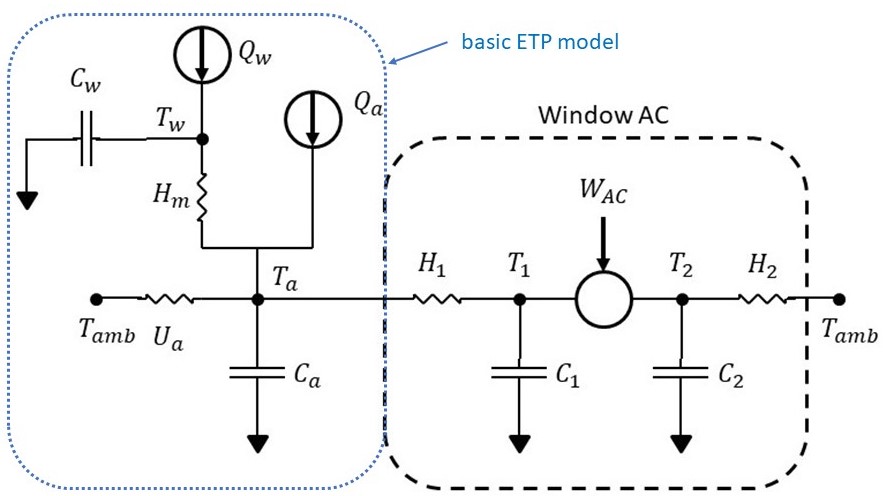}
\caption{The circuit diagram used for a model house.  The basic ETP model from \cite{etp-model} is given in the dotted blue box on the left. In this basic model, $Q_a$ would represent the air conditioner as simply a step function with heating rate $Q_a = 0$ when the AC is off and $Q_a = -Q_c$ when the AC is on.  Here, though, we extended the model to explicitly include a simple model of the air conditioner (black, dashed box) to account for the time-varying power draw in cooling the house. Including this lossy Carnot refrigerator model introduces physics missing from the basic model, such as the time lag for the air conditioner's heat exchangers to warm or cool and the effect of the outside temperature on the real power consumption of the AC (cf. Fig.~\ref{fig:pVsT}a).}
\label{circuit-diagram} 
\end{figure}

The model used here is depicted in Fig.~\ref{circuit-diagram}.  The heat injected into the water is $Q_w$, and the water has a temperature $T_w$ and heat capacity $C_w$.  Generally, the direct heating of air $Q_a$ is negligible compared to the other heat loads since there is always a solid (e.g., windings in the duct fan) that is directly heated and transfers heat to the air from its surface.  In that case, $Q_a \approx 0$ and the heat generation can effectively be grouped with $Q_w$.  The outside temperature $T_{\rm amb}$ affects the system both through heat conducted through the walls, with conductance $U_a$, and also by being the temperature of the reservoir into which the AC rejects heat.

The heat flow equations for this model are:
\begin{eqnarray}
    C_w \dot{T}_w & = & H_m \left( T_a -T_w \right) + \dot{Q}_w + \dot{Q}_a \label {tweqn}\\
    C_r \dot{T}_a & = & U_a \left( T_{\rm amb} - T_a \right) + H_m \left( T_w - T_a \right) \nonumber \\
    & & \;\;\; +\; \dot{Q}_{\rm a} + H_1 \left( T_1 - T_r \right) \\
    C_1 \dot{T}_1 & = & H_1 \left( T_r - T_1 \right) - \dot{Q}_c \\
    C_2 \dot{T}_2 & = & H_2 \left( T_a -T_2 \right) + \dot{Q}_c + \dot{W},
    \label{t2eqn}
\end{eqnarray}
where $T_a$ is the air temperature, $T_1$ is the temperature of AC's cold heat exchanger (evaporator), $T_2$ is the temperature of the AC's hot heat exchanger (condenser), $H_m$ is the heat transfer coefficient from the water to air, $H_1$ and $H_2$ are the heat transfer coefficients of the AC's heat exchangers, the $C_i$ are the heat capacities of the bulk model components, $Q_c$ is the heat removal rate of the air conditioner, and $W$ is the real power consumption of the AC.

As described in Appendix A, we model the compressor performance as
\begin{eqnarray}
    \dot{Q}_c & = & A \exp \left( -L/R T_1 \right) /T_1 \\
\dot{W} & = & \gamma \dot{Q}_c \frac{T_2- T_1}{T_1} + W_{\rm fric}
\end{eqnarray}
where $A$ is a constant defined in the appendix, $L$ is the latent heat of the refrigerant (hydrofluorocarbon mixture R-410A for our window AC unit under test), $R$ is the universal gas constant, $\gamma$ is an arbitrary factor for extra loss in the refrigerant loop (thermal gradients or heat leaks, etc.), and $W_{\rm fric}$ is a constant loss from resistance or friction in the compressor.  These latter relations render the system of equations nonlinear, and so we solve them numerically for the temperatures as functions of the time.  In this work, the model was programmed in Mathematica and the NDSolve function was used to solve the system of equations until the air temperature $T_a$ reached the $T_-$ deadband limit.  Then the equations were solved with $Q_c =0$ until $T_a$ reached the $T_+$ limit.  This was repeated until the piece-wise temperature trajectory reached a steady-state cycling pattern.

\begin{figure}
\centering
\includegraphics[width=8.5 cm]{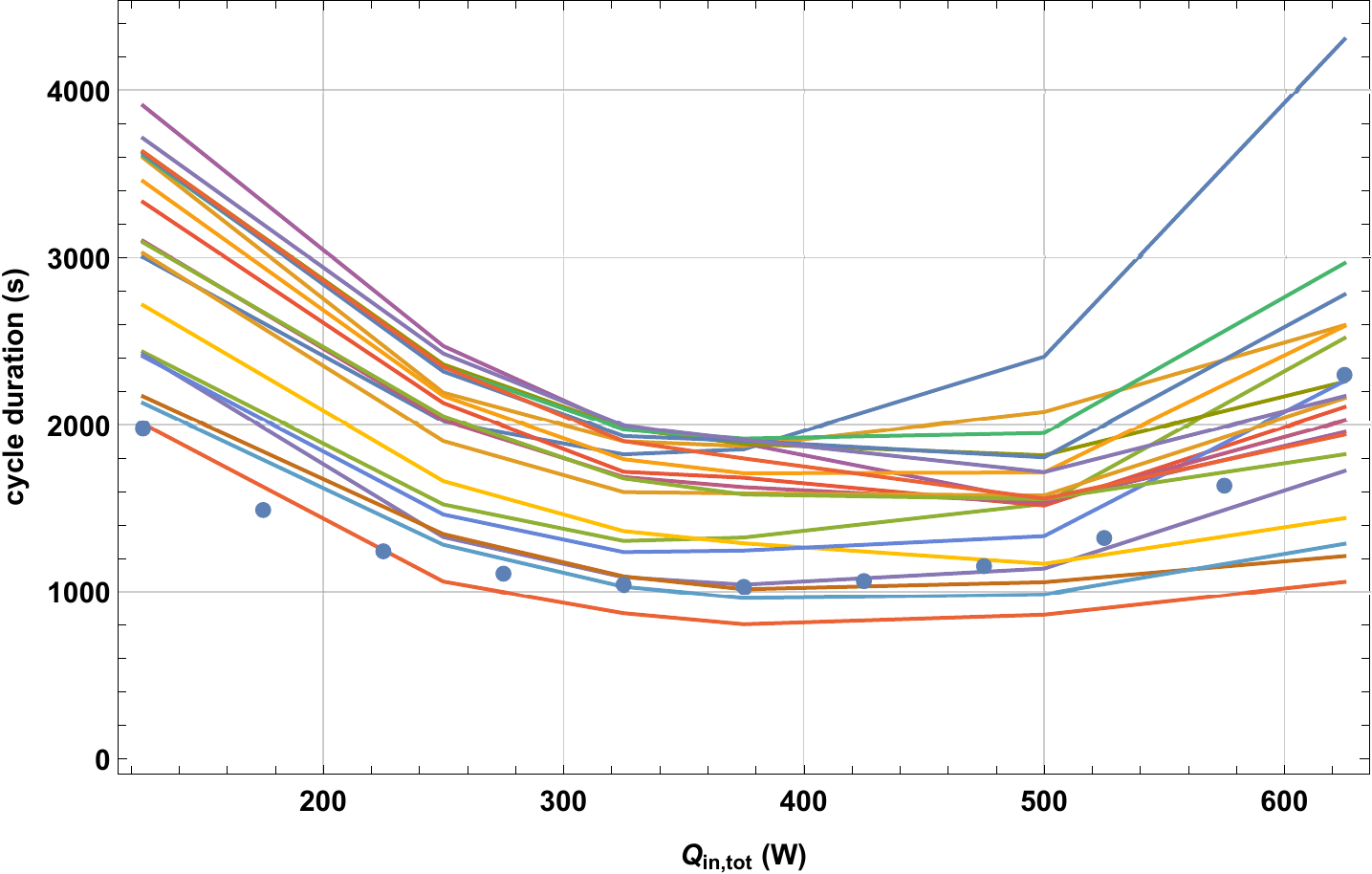}
\caption{Cycle durations vs.~internal heating rate for the physical TCLs (lines) and for the extended ETP model (circles).  The injected heat $Q_{\rm in,tot}$ includes the fixed heating rate of the duct fan and the water pump, which together add 125 W to the nominal programmed value.\label{cycle-durations}}
\end{figure}

For benchmarking the experimental testbed against summertime field test data from Pecan Street, Inc.~\cite{PSI}, we wanted the cycling patterns of the ACs in each testbed, such as the distribution of cycle durations, be similar.  The AC data in the field testbed varied for each house with the outside air temperature, irradiance by the sun, occupancy, and temperature setpoints chosen by homeowners.  Using the ETP model, one can show the variation of the AC's ON-time, OFF-time, and resulting cycle duration as a function of the heat gain in the unit.  The results from this model are plotted against data from the model houses of the experimental testbed in Fig.~\ref{cycle-durations} and are seen to be similar in form.  For low internal heating rates, the air conditioner cools the room quickly, but the temperature rises with time constant $\sim \left( C_a + C_w \right) \left( T_+ - T_- \right)/Q_{\rm in,tot}$, where $Q_{\rm in,tot}$ is the total heat injection from steady and programmable sources.  For heating rates approaching $Q_c$, the maximum cooling rate of the AC, the time constant for cooling diverges with $\left( C_a + C_w \right) \left( C_a + C_w \right) \left( T_+ - T_- \right)/(Q_c- Q_{\rm in,tot})$ as $Q_{\rm in,tot} \rightarrow Q_c$.  The cycle duration is therefore lowest for moderate ($\sim 50\%$) duty cycle of the compressor, but the range of heat injections producing low durations is quite broad.

Although the twenty houses are constructed to be nearly identical, the heating and cooling rates and resulting cycle durations were found to span a wide range in length.  The units with the shortest cycles were found to have higher time-averaged water temperatures.  This might suggest that one should improve the thermal contact between air and water, so that more of the water's heat capacity is effectively accessed in each thermal cycle.  One way to do this would be to increase the size of the heat exchanger, e.g., by increasing the size of the radiator in the hydronic system.  However, this requires altering the design and should not be necessary given that other identical units operate with useful cycle times, i.e., ON and OFF times that are longer than the compressor's lockout time.  The AC's recommended lockout time varies across different models and sizes, but 3 to 5 minutes is typical.  For the ACs used in this testbed, the manufacturer specified 3 minutes.

Increasing the average air flow speed through the heat exchanger is \emph{not} a solution that increases the cycle duration in a model house.  Assuming fully-developed laminar air flow in the heat exchanger, the convection coefficient $h$ is nearly constant with air speed (see (8.53) of \cite{incropera}), so that the instantaneous heat transfer rate
\begin{equation}
   \dot{q}_{HX} = h \left( T_{w} - T_{a} \right), \label{qOfHdT}
\end{equation}
would be unchanged.  In these units, the air flow is fast enough that it does {\em not} reach fully-developed thermal or velocity profiles within the depth of the heat exchanger, even though the flow is still laminar.  The ``entrance region" spans the whole thickness of the heat exchanger, and $h$ does increase with air speed.  Still, the average heat transfer coefficient is only increasing with the cube root of the velocity from its value for fully-developed laminar flow until one reaches the transition to turbulent flow~\cite{incropera}.  In spite of the increase in heat transfer rate, the thermometer used for the thermostat will see a decrease in its coupling to the water temperature.  The thermometer is immersed in the exit flow from the heat exchanger, and an increase in air speed through it results in a lower temperature difference relative to the well-mixed air in the model house, because
\begin{equation}
    \dot{q}_{HX} = \rho c_p v \left( T_{out} - T_{in} \right), \label{qOfRhoCVdT}
\end{equation}
where  $\rho$ is the density of air, $c_p$ is the specific heat of air, $v$ is the air speed, and $T_{\rm out}$ and $T_{\rm in}$ are the temperatures of the air entering and leaving the heat exchanger.  Even if $\dot{q}_{HX}$ is increased slightly, the linear change in $v$ ensures that the thermostat is brought closer to the true, mixed temperature of the air in the box,
not to the temperature of the water.

To utilize the heat capacity of the water and slow down thermal cycling, one needs instead to reduce the air flow speed around the thermometer so that it is more strongly coupled to the water temperature.  No flow straighteners and buffer ducts were used at the inlet to the heat exchanger, and the fan is round while the heat exchanger is square.  In that case, the air flow profile across the exchanger is highly inhomogeneous and one can place the thermometer in different air flows simply by repositioning it.  A vane anemometer can be used to measure the air flow speed at any location on the heat exchanger's face.  Changing the air speed at the thermometer in this way does not affect the convection coefficient of the heat exchanger at all.

\begin{figure}
    \centering
    \includegraphics[width=8.5 cm]{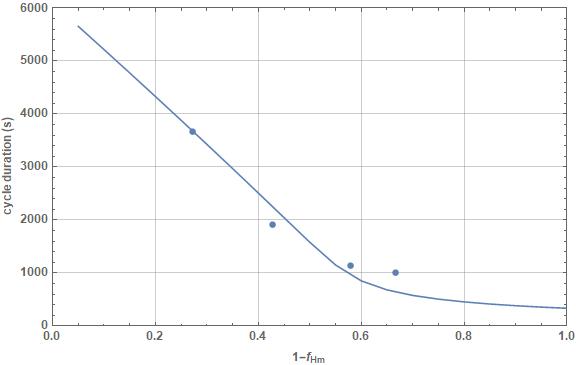}
    \caption{Cycle duration vs.\ thermometer placement in an ETP model is shown as the solid curve.  For $1-f_{Hm}$ approaching 1, the thermostat is using the mixed air temperature in the middle of the box, and the air temperature is weakly coupled to the water so that the TCL cycle durations are fastest.  As $1-f_{Hm}$ approaches 0, though, the thermostat is tightly coupled to the water temperature, and the entire heat capacity of the water must heat or cool to the temperature limits in each cycle.  Circles are data from a physical model house run with its thermostat sensor at four different positions above the air-water heat exchanger.  Because $h$ in \eqref{eqn:feff} is not measured, the data are scaled to the curve at a single point. The model was not calibrated to this particular model house, but the trend is similar.}
    \label{fig:fHm_plots}
\end{figure}

Relocation of the thermostat's thermometer can be explored in the ETP model above by defining a new variable 
\begin{equation}
    T_{\rm therm} \equiv (1 -f_{Hm}) T_{a} + f_{Hm} T_{w},
\end{equation}
where $0 \leq f_{Hm} \leq 1$ is a parameter chosen to represent how close the thermometer is to the well-mixed air temperature versus the water temperature.  When $f_{Hm} = 0$, $T_{\rm therm}=T_{a}$ as before; but when $f_{Hm} = 1$, $T_{\rm therm}=T_{w}$ as if the thermometer is immersed in the water.  This equation is evaluated along with the heat flow equations \eqref{tweqn}-\eqref{t2eqn}, and $T_{\rm \rm therm}$ is then used for the thermostat to determine when the $T_-$ and $T_+$ deadband limits are reached in the numerical integration.  Because most of the heat injection is through the water heater, the air temperature $T_{a}$ generally does not overshoot $T_{w}$ or $T_{\rm therm}$.  However, $T_{a}$ typically is much lower than $T_{\rm therm}$ during a cooling cycle as the AC cools the air directly.  In Fig.~\ref{fig:fHm_plots} the full cycle duration for the model is plotted against $1-f_{Hm}$, the relative coupling of the thermostat to the true air temperature.  Subsequently, experiments were performed on one of the model houses in which the thermostat's thermometer was positioned at several different locations above the air-water heat exchanger, and the air flow at those locations was measured with an anemometer.  To relate the air speed to an effective $1-f_{Hm}$, we can use \eqref{qOfHdT} and \eqref{qOfRhoCVdT}, with $T_{out} \equiv T_{\rm therm}$ and $T_{in} \equiv T_{a}$ to find
\begin{equation}
    f_{\rm eff} = \frac{h}{\rho c_p}. \label{eqn:feff}
\end{equation}
The data are scaled for comparison with the qualitative simulation in Fig.~\ref{fig:fHm_plots} as well.

\section{Discussion} \label{all-testbed}
\begin{figure}
\centering
\includegraphics[width=8.5 cm]{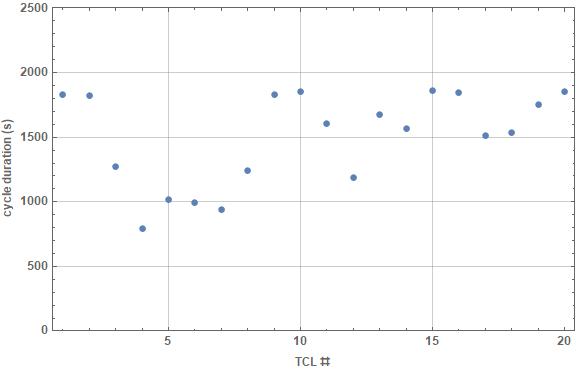}
\caption{Cycle durations vs. index for particular thermostat and heating rate settings.\label{cycle-durations-ensemble}}
\end{figure}

After construction, the 20-unit testbed was exercised over a range of settings, varying the temperature setpoint, the temperature deadband, and the heating rate injected into the water.  For fixed settings, the testbed was allowed to run for several hours, with the first two or three hours needed for the testbed to reach a quasi-steady state in which the water and air temperatures were cycling repeatedly between the same limits.  The heterogeneity of the houses can be seen in Fig.~\ref{cycle-durations-ensemble} showing the distribution of cycle durations of the houses all under the same thermostat and heating conditions.  The outlier TCLs with cycle durations close to the ``lockout" time for preventing short-cycling were adjusted by moving their temperature sensors as described in Section~\ref{ETPmodeling}.  
Further tuning could have been done on other TCLs in the ensemble to make the overall distribution more homogeneous.  This could have been used to cause artificially coherent power fluctuations, as groups switched on at the same time would continue to turn on simultaneously in later cycles.  However the goal of this project was to look for fluctuations in a realistic distribution similar to field tests.

In spite of the close proximity of units to one another, there was no phase-locking or coincidence of transitions seen in neighboring units, so they acted like independent oscillators.  

Additional testing was done with experiments that sent regularly spaced ON/OFF requests to all units.  
The spacing of these square-waves of ON and OFF requests were varied to look for long-time synchronization of the ACs.  Because the ACs are independent and have different natural cycling periods for fixed conditions, though, no strong synchronization was seen.  When the majority of ACs are prepared into the ON or OFF states by a series of state requests, releasing the ensemble by discontinuing control leads to a rapid de-phasing of the cycling of the ACs, as expected.

\section{Conclusions} \label{conclusion-section}
We have constructed a system with 20 physical model houses for tests of demand-response control methods. The physical models provide data on power usage and temperature response that can be used to create or improve high-fidelity models of houses for use in simulations. We have shown that the physical models can be tuned in cycle duration by careful placement of their thermometers.

Having benchmarked the testbed, experiments are underway to test control methods, such as model-based predictive controllers with feedback.  The purpose is to determine what types of controllers provide faithful tracking of the control signal and to try to identify conditions that could cause control to degrade.  The ability to adjust the natural cycling of the TCLs allows for tests on more homogeneous ensembles too, such as might be found in a block of small offices.

\section*{APPENDIX}
This section describes the simple model of an AC used to extend our ETP model.  This model contributes a time-varying real power consumption when the AC is turned on.
The compressor in the air conditioner works at a single speed and has only the binary ON/OFF states.  When ON, then, the compressor has a fixed rate of displacement, $\dot{V}$, for the working fluid, refrigerant R410a.  The mass flow then depends on the density of the refrigerant at the inlet to the compressor, or
\begin{equation}
    \dot{m}=\rho_c(T) \;\dot{V},
    \label{mdoteqn}
\end{equation}
where the density $\rho_c$ is a function of the temperature $T$ of the cold heat exchanger, from which the heat is being pumped.

Although the refrigerant is cold, because it is the material evaporated from the cold heat exchanger, we approximate the vapor as an ideal gas so that
\begin{equation}
    \rho_c = \frac{P_c}{R T}
\end{equation}
where $R=R_{\rm univ}/M_{\rm mol}$ is the molecular weight-dependent gas constant for this material.  The vapor pressure for the fluid may be approximated as
\begin{equation}
    P_c \propto \exp\left( -L/RT \right)
\label{vaporpressurecurve}
\end{equation}
where $L$ is the latent heat for the liquid-vapor transition.  This latent heat is typically a weak function of temperature, so it is taken to be constant in this approximate model.  The value for latent heat in this model was chosen such that the simplified formula~\eqref{vaporpressurecurve} best matches the vapor pressure curve for the refrigerant around room temperature, up to an amplitude prefactor.

Finally, the cooling power of the compressor loop will be
\begin{equation}
    \dot{Q}_c = \dot{m} \; L
    \label{coolingpowereqn}
\end{equation}
at the cold heat exchanger.  Putting together expressions \eqref{mdoteqn}--\eqref{coolingpowereqn} leads to
\begin{equation}
    \dot{Q}_c = A \frac{\exp\left( -L/RT \right)}{ T }
\end{equation}
where $A$ includes $\dot{V}$, $R$, and the proportionality constant in \eqref{vaporpressurecurve}.

\section*{ACKNOWLEDGMENT}
The authors are grateful to Scott Backhaus for the original concepts and motivation in developing these model systems.  We also thank Greg Swift for many useful discussions, and our ARPA-E project collaborators: Duncan Callaway, Ioannis Granitsas, Scott Hinson, Ian Hiskens, and Oluwagbemileke Oyefeso.  

\bibliographystyle{IEEEtran}	
\bibliography{testbed}

\end{document}